\documentclass[12pt]{article}
\usepackage{epsfig}
\usepackage{amsmath}
\usepackage{hhline}
\usepackage{amssymb}
\usepackage{times}
\usepackage{color}

\newlength{\dinwidth}
\newlength{\dinmargin}
\newcommand{\picob}{\mbox{{\rm ~pb}}}

\def\GeV{\hbox{$\;\hbox{\rm GeV}$}}

\begin{document}

\pagestyle{empty}
\begin{titlepage}
\noindent

\begin{flushleft}
DESY 05-087 \hfill ISSN 0418-9833 \\
June 2005
\end{flushleft}

\vspace{2cm}

\vspace{2cm}

\begin{center}
  \Large
  {\bf 
   Search for Leptoquark Bosons in {\boldmath{$ep$}} Collisions at HERA}

  \vspace*{1cm}
    {\Large H1 Collaboration} 
\end{center}

\begin{abstract}
\noindent
A search for scalar and vector leptoquarks coupling to
first generation fermions is performed 
using the $e^+p$ and $e^-p$ scattering data collected by the H1 experiment
between 1994 and 2000. 
The data correspond to a total integrated luminosity of $117$\,pb$^{-1}$.
No evidence for the direct or indirect production of such
particles is found in data samples with a large transverse
momentum final state electron or with large missing transverse
momentum.
Constraints on leptoquark models are established. For
leptoquark couplings of electromagnetic strength,
leptoquarks with masses up to $275-325\,{\rm GeV}$ are ruled out.
These limits improve and supercede earlier H1 limits based on 
subsamples of the data used here.

\noindent
\end{abstract}

\vspace{5mm}
\begin{center}
(To be submitted to Phys.\ Lett.\ B)
\end{center}

\end{titlepage}

\begin{flushleft}

A.~Aktas$^{10}$,               
V.~Andreev$^{26}$,             
T.~Anthonis$^{4}$,             
S.~Aplin$^{10}$,               
A.~Asmone$^{34}$,              
A.~Astvatsatourov$^{4}$,       
A.~Babaev$^{25}$,              
S.~Backovic$^{31}$,            
J.~B\"ahr$^{39}$,              
A.~Baghdasaryan$^{38}$
% ...

\bigskip{\it
 $ ^{1}$ I. Physikalisches Institut der RWTH, Aachen, Germany$^{ a}$ \\
% ...
$ ^{4}$ Inter-University Institute for High Energies ULB-VUB, Brussels;
          Universiteit Antwerpen, Antwerpen; Belgium$^{ c}$ \\
% ...
$ ^{10}$ DESY, Hamburg, Germany \\
% ...
$ ^{25}$ Institute for Theoretical and Experimental Physics,
          Moscow, Russia$^{ l}$ \\
 $ ^{26}$ Lebedev Physical Institute, Moscow, Russia$^{ e}$ \\
% ...
$ ^{31}$ Faculty of Science, University of Montenegro,
          Podgorica, Serbia and Montenegro$^{e}$ \\
% ...
$ ^{34}$ Dipartimento di Fisica Universit\`a di Roma Tre
          and INFN Roma~3, Roma, Italy \\
% ...
$ ^{38}$ Yerevan Physics Institute, Yerevan, Armenia \\
 $ ^{39}$ DESY, Zeuthen, Germany \\
% ...
$ ^a$ Supported by the Bundesministerium f\"ur Bildung und Forschung, FRG,
      under contract numbers 05 H1 1GUA /1, 05 H1 1PAA /1, 05 H1 1PAB /9,
      05 H1 1PEA /6, 05 H1 1VHA /7 and 05 H1 1VHB /5 \\
% ...
$ ^c$ Supported by FNRS-FWO-Vlaanderen, IISN-IIKW and IWT
      and  by Interuniversity
Attraction Poles Programme,
      Belgian Science Policy \\
% ...
$ ^e$ Supported by the Deutsche Forschungsgemeinschaft \\
% ...
$ ^l$ Partially Supported by Russian Foundation
      for Basic Research, grant    no. 00-15-96584 \\
}
\end{flushleft}

\newpage
\pagestyle{plain}

\section*{Introduction}

The $ep$ collider HERA offers the unique possibility to search for the 
resonant production of new particles which couple directly to
a lepton and a parton.
Leptoquarks (LQs), colour triplet bosons which appear naturally
in various unifying theories beyond the Standard Model (SM), 
are such an example.
At HERA, leptoquarks could be singly produced by the fusion of the
initial state lepton of energy $27.6 \GeV$ with a quark from the
incoming proton of energy up to $920 \GeV$.

The phenomenology of LQs at HERA is discussed in detail in~\cite{H1LQ99}.
The effective Lagrangian considered conserves lepton and baryon number,
obeys the symmetries of the SM gauge groups ${\rm U}(1)_Y$, ${\rm SU}(2)_L$ 
and ${\rm SU}(3)_C$
and includes both scalar and vector LQs.
A dimensionless parameter $\lambda$ defines the coupling at 
the lepton-quark-${\rm LQ}$ vertex.
At HERA, LQs can be resonantly produced in
the $s$-channel or exchanged in the $u$-channel between the
incoming lepton and a quark coming from the proton.
In the $s$-channel, a LQ is produced at a mass $M =\sqrt{xs_{ep}}$
where $x$ is the momentum fraction of the proton carried by the 
interacting quark.

This letter presents a search for LQs coupling to first generation 
fermions using $e^+ p$ data collected at a centre-of-mass energy of
$\sqrt{s_{ep}} \approx 300 \GeV$, 
$e^- p$ data collected at $\sqrt{s_{ep}} \approx 320 \GeV$
and $e^+ p$ data collected at $\sqrt{s_{ep}} \approx 320 \GeV$.
These data sets correspond to integrated luminosities of
$37 \picob^{-1}$, $15 \picob^{-1}$ and $65 \picob^{-1}$ respectively.
They represent the full statistics accumulated by the H1 experiment 
between 1994 and 2000. The results of this search thus supercede those
based on the $e^+p$ 1994-1997~\cite{H1LQ99} 
and $e^-p$ 1998-1999~\cite{H1LQe-} data.

The $e^+ p$ and $e^- p$ data are largely complementary
when searching for leptoquark resonances.
Due to the more favourable parton densities of quarks with respect to
anti-quarks at high $x$, the $e^{+}p$ ($e^{-}p$) data sets are mostly
sensitive to LQs with fermion number\footnote{
 The fermion number $F$ is given by $F=3B+L$ with $B$ and $L$ being
 the baryon and lepton numbers respectively.}
$F=0$ ($F=2$).
The search reported here considers the leptoquark decays into
an electron and a quark and into a neutrino and a quark.
These LQ decays lead to final states similar
to those of deep-inelastic scattering (DIS) neutral current (NC)
and charged current (CC) interactions at very high $Q^2$,
the negative four-momentum transfer squared. 
If the final state consists of an electron and a quark,
the LQ mass is reconstructed from the measured
kinematics of the scattered electron. 
If the final state consists of a neutrino and a quark,
the LQ mass is
reconstructed from the hadronic final state~\cite{H1LQ99}.

\section*{Neutral and Charged Current Data}

The H1 detector components most relevant to this analysis are the liquid argon
calorimeter, which measures the positions and energies of
charged and neutral particles over the range 
$4^\circ<\theta<154^\circ$ of polar angle\footnote{The polar 
angle $\theta$ is defined with respect to the incident proton momentum vector 
(the positive $z$ axis).},
and the inner tracking detectors which measure
the angles and momenta of charged particles over the range
$7^\circ<\theta<165^\circ$. A full description of the detector can be
found in~\cite{h1det}. 

This search is based on inclusive NC and CC DIS data in the kinematic
domain $Q^2>2500\,\mathrm{GeV}^2$ and $0.1<y<0.9$, where 
the inelasticity variable $y$ is defined as $y=Q^2/M^2$.
The cuts on $y$ remove regions of poor reconstruction, poor resolution,
large QED radiative effects and background from photoproduction
processes. 
The selection of NC-like events follows that presented
in~\cite{H1LQ99}.
It requires an identified electron with transverse momentum above
$15 \GeV$.
The selection of CC-like events follows closely that
presented in~\cite{H1LQ99,H1EMINUS}.
A missing transverse momentum exceeding $25 \GeV$ is required.

The inelasticity variable $y$ is related to
the polar angle $\theta^\ast$ of the decay lepton
in the centre-of-mass frame of the hard subprocess $(eq\rightarrow lq)$
by $y =\frac{1}{2}(1+\cos\theta^\ast)$.
Since the angular distribution of the electron coming from the decay
of a scalar resonance is markedly different from that
of the scattered lepton in NC DIS~\cite{H1LQ99}, a mass
dependent cut on $y$ was applied previously~\cite{H1LQ99,H1LQe-}
in order to optimise the signal sensitivity. 
However, the optimisation power is rather limited 
for a vector resonance as the angular distribution is only slightly
different from that of the DIS background. 
For this reason, no such mass dependent $y$ cut is applied in this
analysis.
Instead, all selected events are analysed in bins of varying size adapted to 
the experimental resolution in the $M-y$ plane, with a procedure
designed to fully exploit the sensitivity to the signal as explained 
in the next section.

\begin{figure}[p] 
\begin{center}
\begin{picture}(50,155)
\put(-47.5,-5){\epsfig{figure=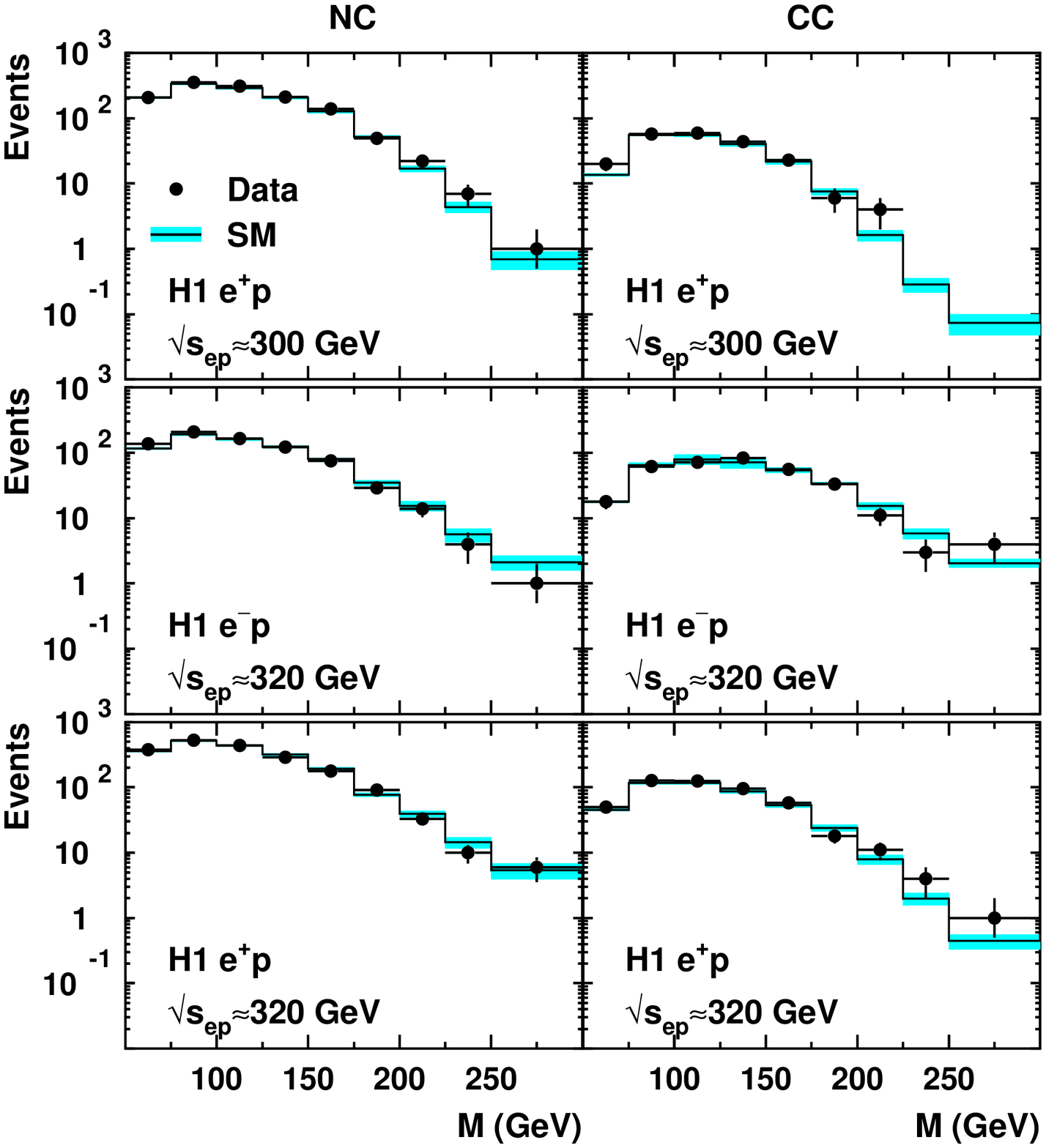,width=14cm}}
\put(20,130){\bf (a)}
\put(20,85){\bf (b)}
\put(20,45){\bf (c)}
\put(77,130){\bf (d)}
\put(77,85){\bf (e)}
\put(77,45){\bf (f)}
\end{picture}
\end{center}
  \caption{\label{fig:dndm}
  Mass spectra for the 
  (a-c) neutral current (NC) and (d-f) charged current (CC) deep inelastic 
  scattering selected events, together with the corresponding Standard
  Model (SM) expectations.
  The shaded bands indicate the $\pm1\sigma$ uncertainty 
  on the SM expectations.}
\end{figure}

The mass spectra measured for NC- and CC-like events in the three data sets
are compared in Figs.~\ref{fig:dndm}a-f
with the SM predictions, obtained using a Monte-Carlo (MC)
calculation~\cite{DJANGO} and the CTEQ5D parametrisation~\cite{cteq5d}
for the parton densities. In all cases the data are well described by
the SM prediction.
Since no evidence for LQ production is observed in either the NC or CC
data samples, the data are used to set constraints on LQs which couple
to first generation fermions.

\section*{Statistical Method}

\begin{figure}[p] 
\begin{center}
\begin{picture}(50,155)
\put(-47.5,-5){\epsfig{figure=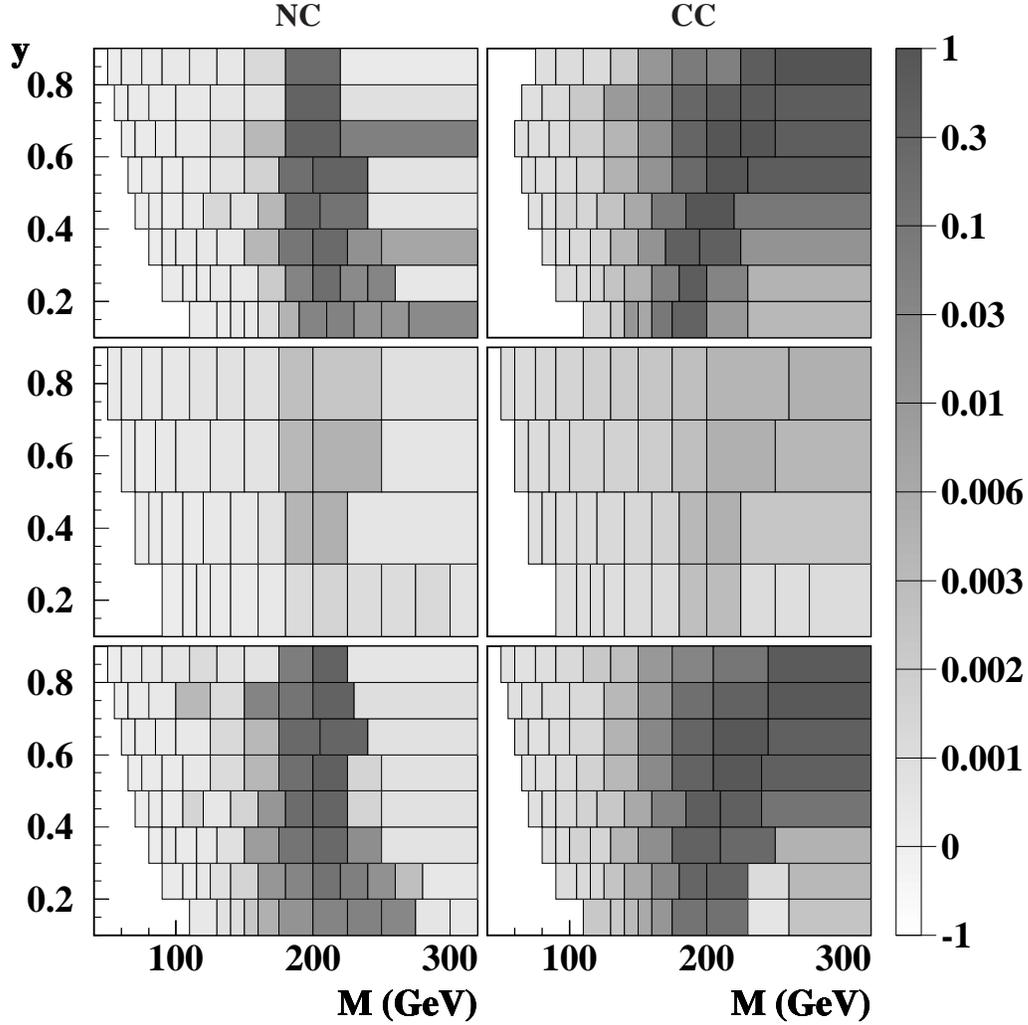,width=14cm}}
\put(-10,128){\bf NC}
\put(42.5,128){\bf CC}
\end{picture}
\end{center}
  \caption{\label{weight_bin}
  Binning used in the $M-y$ plane for the different data sets and
  weights calculated in these bins for a 200 GeV vector 
  leptoquark with a coupling of $0.023$ to $e^{+}d$ and $\bar{\nu}u$.
  The left plots correspond to the neutral current (NC)-like decay channel 
  whereas the right plots correspond to the charged current (CC)-like decay 
  channel. 
  The top, middle and bottom plots correspond to the $300$\,GeV $e^+p$,
  $320$\,GeV $e^-p$ and $320$\,GeV $e^+p$ data sets, respectively.}
\end{figure}

For the limit analysis, the data are studied in bins in 
the $M-y$ plane. The binning used is
different for the different data sets and is shown in Fig.~\ref{weight_bin}.
For those LQ types with only a NC-like
decay mode, the total number of bins amounts to about $200$ covering 
all three NC samples.
The total number of bins doubles when including the three CC data samples
for LQs having both NC-like and CC-like decay modes.
The number of SM background events $b_i$ in each bin $i$ 
is obtained from the SM MC calculations. 
Each MC event $k$, reconstructed in bin $i$, has an event weight 
$e_k$, such that the MC is normalised to the luminosity of the data. The
sum over all SM MC events within bin $i$ thus gives
\begin{equation}
b_i=\sum_{k\in {\rm bin}\, i} e_k\,.
\end{equation}
To estimate the LQ signal, an event re-weighting technique is applied to the
same SM MC events. No use is made of a dedicated signal MC generator.
The number of events expected in each bin $i$ 
in the presence of a LQ signal is denoted as $s_i+b_i$. It may be written as
\begin{equation}
s_i+b_i=\sum_{k\in {\rm bin}\, i} e_k\frac{\sigma^{\rm LQ}_k + 
                                         \sigma^{\rm INT}_k +
                                         \sigma^{\rm SM}_k}
                                        {\sigma^{\rm SM}_k}\,,
\label{eqn:bsweight}
\end{equation}
where $\sigma^{\rm LQ, INT, SM}_k$ are differential cross section
terms~\cite{BRW} corresponding to the LQ, interference and SM contributions, respectively.
These differential cross section terms,
calculated in leading order using the parton density functions 
CTEQ5D~\cite{cteq5d}, are based on the true kinematic 
quantities of event $k$, whereas the resulting $s_i+b_i$ events are
counted in the $M-y$ bins of the reconstructed variables with appropriate
simulation of the detector response.
The differential cross section terms $\sigma^{\rm LQ, INT}_k$ depend on the LQ mass 
and coupling $\lambda$. For mass values well below 
the kinematic limit $\sqrt{s_{ep}}$, the $s$-channel contribution dominates 
in the LQ cross section and the signal contribution $s_i$ to bin $i$
is always positive. 
However, at higher masses, the interference contributions become more important
and $s_i$ may be negative in the case of destructive interferences, although
$s_i+b_i$ always stays positive.

The limits are determined from a statistical analysis which uses the 
method of fractional event counting~\cite{bock}. 
For a given leptoquark mass and coupling 
a weight $w_i$ is ascribed to each bin, which is given by
the asymmetry between the expected number of events in the presence
or absence of a LQ signal:
\begin{equation}
w_i = \frac{(s_i+b_i)-b_i}{(s_i+b_i)+b_i}=\frac{s_i}{s_i+2b_i}\,.
\end{equation} 

\noindent As an example, Fig.~\ref{weight_bin} illustrates the bin dependent weights for 
an $e^+d$ type vector LQ having a mass of $200$\,GeV,
a coupling of $0.023$ and both the NC-like and CC-like decay channels.
As expected, the weights are small for the $e^-p$ data.
For the $e^+p$ data sets, the weights have little $y$-dependence.

Using these weights a fractional event count, also called test
statistics, is defined as
\begin{eqnarray}
\label{X_data}
   X({\rm data})&=&\sum_i w_i N_i({\rm data})\,,
\end{eqnarray}
where $N_i({\rm data})$ is the number of data events observed in bin $i$.

In a frequentist approach, a large number of ``experiments'' ($2\times
10\,000$) are 
generated. Each experiment consists of Poisson distributed random
numbers $N_i(b)$ $(N_i(s+b))$, based on the expected number of
events $b_i$ $(s_i+b_i)$ in the absence (presence) of a
LQ signal. For each background experiment $b$ and for each
signal-plus-background experiment $s+b$ a fractional event count is defined in
analogy to Eqn.(\ref{X_data}):
\begin{eqnarray}
   X(b)&=&\sum_i w_i N_i(b)\\
   X(s+b)&=&\sum_i w_i N_i(s+b)\,.
\end{eqnarray}
Frequentist probabilities ${\rm CL}_{s+b}$ $({\rm CL}_b)$ 
are defined as the fraction of experiments where the quantity
$X(s+b)$ $(X(b))$ is smaller than $X({\rm data})$.
If the data agreed perfectly with the expectation from the background-only
hypothesis, a value of ${\rm CL}_b=0.5$ would be obtained.
A higher value indicates that the observation is more signal-like; 
a lower value indicates fluctuations opposite to those expected for a signal.
If ${\rm CL}_{s+b}$ is small, it may be used to exclude
the signal-plus-background hypothesis with confidence level 
($1-{\rm CL}_{s+b}$). However, in this analysis we use the confidence level 
${\rm CL}$ defined as
\begin{equation}
{\rm CL} = 1-{\rm CL}_{s+b} / {\rm CL}_b
\end{equation}
to set limits in a conservative manner. This ratio, which was also used in LEP
searches~\cite{bock, opallq}, has the desirable feature that as the LQ
coupling  tends to zero, and necessarily 
${\rm CL}_{s+b} \rightarrow {\rm CL}_b$, 
${\rm CL}$ drops to $0$, i.e. one cannot rule out any LQ which has
a vanishing coupling.
On the other hand, for a non-zero coupling, an exclusion limit at 
$95\%\,{\rm CL}$ is always reachable for a sufficiently large coupling.

Systematic uncertainties enter as offsets $\delta^{b}_{i,j}$
and $\delta^{s+b}_{i,j}$ to the predicted number of events $b_i$ and
$s_i+b_i$, where $j$ runs over all independent sources of systematic
errors. For the limits presented below, both the bin weights $w_i$ and 
the $b$ and $s+b$ MC experiments are altered by the known systematic 
uncertainties, assuming that the latter have Gaussian probability densities.

The experimental systematic error is dominated by the electromagnetic energy
scale (between $0.7 \%$ and $3 \%$) for the NC analysis, and by
the hadronic energy scale ($2 \%$)
for the CC analysis.
The limited knowledge of proton structure causes an uncertainty on the
signal cross section. 
This uncertainty is estimated to be $5 \%$ for $F=2$ ($F=0$) LQs coupling to 
$e^- u$ ($e^+ u$) and varies between $7 \%$
at low LQ masses up to $30 \%$ around 290\,GeV for 
$F=2$ ($F=0$) LQs coupling to $e^- d$ ($e^+ d$). Similarly, an
uncertainty on the DIS cross sections is connected with the parton densities.
The correlation between the systematic uncertainties on the signal and 
that on the background and between different analysis bins is taken
into account. 
The correlations induced by the uncertainties of the parton densities are
evaluated using~\cite{botje}.

\section*{Limit Results}

In the following limits will first be derived within
the phenomenological model proposed by
Buchm\"uller, R\"uckl and Wyler (BRW)~\cite{BRW} and then within generic
models where the branching ratios $\beta_e$ ($\beta_{\nu}$)
for the LQ decays into $e q$ ($\nu q$) are not fixed.

The BRW model describes
7 LQs with $F=0$ and 7 LQs with $F=2$.
We use here the nomenclature of~\cite{LQNAME}
to label the various scalar $S_{I,L}$
($\tilde{S}^{\mbox{\tiny
\hspace{-3mm}\raisebox{1.5mm}{(}\hspace{2mm}\raisebox{1.5mm}{)}}}_{I,R}$)
or vector $\tilde{V}^{\mbox{\tiny
\hspace{-3mm}\raisebox{1.5mm}{(}\hspace{2mm}\raisebox{1.5mm}{)}}}_{I,L}$ 
($V_{I,R}$) LQ types of weak isospin $I$, which
couple to a left-handed (right-handed) electron. The tilde is used to
distinguish LQs which differ only by their hypercharge.
In the BRW model the branching ratios $\beta_e$ ($\beta_{\nu}$)
are fixed and equal to 1 or 0.5 (0 or 0.5) depending on the LQ quantum numbers.
Table~\ref{tab:lqbrw} lists the 14 LQ types according to the
BRW model.

\begin{figure}[p] 
\begin{center}
\begin{picture}(50,105)
\put(-47.5,-5){\epsfig{figure=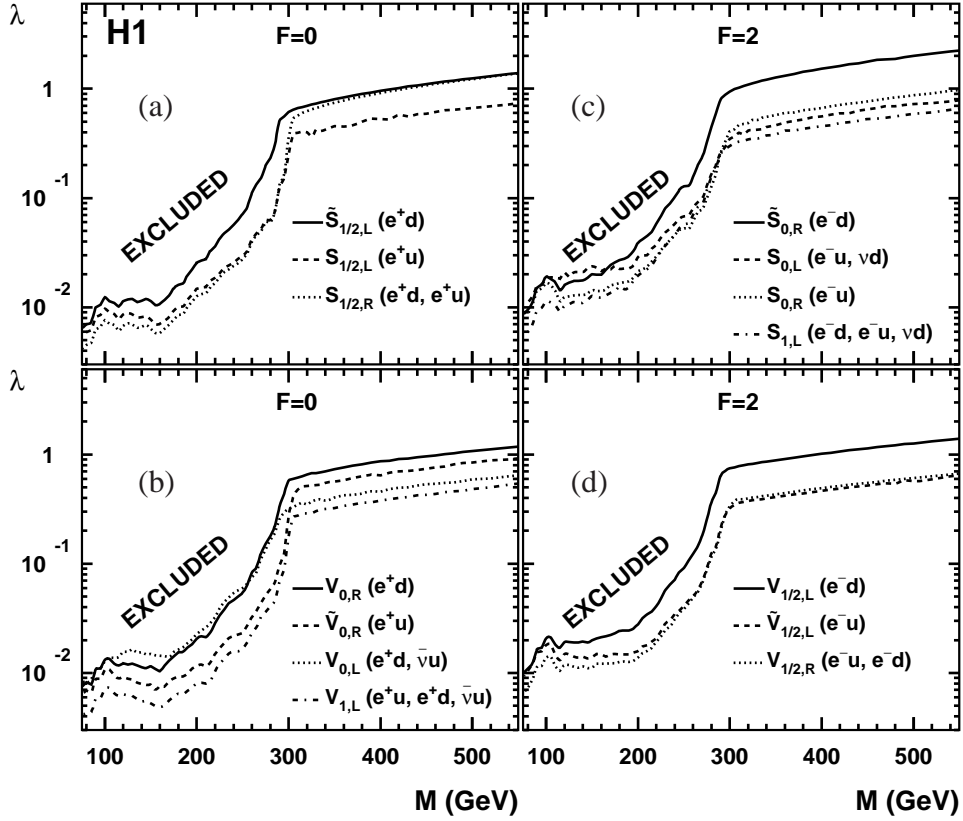,width=14cm}}
\put(-20,90){(a)}
\put(37.5,90){(c)}
\put(-20,40){(b)}
\put(37.5,40){(d)}
\end{picture}
\end{center}
  \caption{\label{fig:brw}
  Exclusion limits for the 14 leptoquarks (LQs) described by the
  Buchm\"uller, R\"uckl and Wyler (BRW) model. The limits are expressed at
  $95\%\,{\rm CL}$ on the coupling $\lambda$ as a function of 
  the leptoquark mass for the (a) scalar LQs with $F=0$, 
  (b) vector LQs with $F=0$, (c) scalar LQs with $F=2$ and 
  (d) vector LQs with $F=2$.
  Domains above the curves are excluded. For each LQ type the pairs of
  Standard Model fermions coupling to it are indicated in brackets (charge
  conjugate states are not shown).}
  \end{figure}

\begin{table*}[htb]
  \renewcommand{\doublerulesep}{0.4pt}
  \renewcommand{\arraystretch}{1.2}
 \vspace{-0.1cm}

\begin{center}
    \begin{tabular}{|c|r|c||c|r|c|}
      \hline
       $F=2$ & Prod./Decay & $\beta_e$
              & $F=0$ & Prod./Decay & $\beta_e$  \\

      \hline
%
     \multicolumn{6}{|c|}{Scalar Leptoquarks} \\ \hline
    $S_{0,L}$     & $e^-_L u_L \rightarrow e^- u$ & $1/2$
  & $S_{1/2,L}$   & $e^+_R u_R \rightarrow e^+ u$ & $1$  \\
                  &           $\rightarrow \nu d$ & $1/2$ & & & \\ \hline
    $S_{0,R}$     & $e^-_R u_R \rightarrow e^- u$ & $1$
  & $S_{1/2,R}$   & $e^+_L u_L \rightarrow e^+ u$ & $1$ \\
      \cline{1-3}
    $\tilde{S}_{0,R}$
                  & $e^-_R d_R \rightarrow e^- d$ & $1$
  &               & $e^+_L d_L \rightarrow e^+ d$ & $1$ \\
      \hline
    $S_{1,L}$     & $e^-_L d_L \rightarrow e^- d$ & $1$
  & $\tilde{S}_{1/2,L}$ 
                  & $e^+_R d_R \rightarrow e^+ d$ & $1$ \\
                  & $e^-_L u_L \rightarrow e^- u$ & $1/2$ & & &  \\
                  &           $\rightarrow \nu d$ & $1/2$ & & & \\
      \hline
%
     \multicolumn{6}{|c|}{Vector Leptoquarks} \\ \hline
    $V_{1/2,R}$   & $e^-_R d_L \rightarrow e^- d$ & $1$
  & $V_{0,R}$     & $e^+_L d_R \rightarrow e^+ d$ & $1$ \\
      \cline{4-6}
                  & $e^-_R u_L \rightarrow e^- u$ & $1$
  & $V_{0,L}$     & $e^+_R d_L \rightarrow e^+ d$ & $1/2$ \\
            & & & &           $\rightarrow \overline{\nu}u$ & $1/2$ \\ \hline
    $V_{1/2,L}$   & $e^-_L d_R \rightarrow e^- d$ & $1$
  & $\tilde{V}_{0,R}$
                  & $e^+_L u_R \rightarrow e^+ u$ & $1$ \\
      \hline
    $\tilde{V}_{1/2,L}$
                  & $e^-_L u_R \rightarrow e^- u$ & $1$
  & $V_{1,L}$     & $e^+_R u_L \rightarrow e^+ u$ & $1$ \\
                          &                                            &
  &               & $e^+_R d_L \rightarrow e^+ d$ & $1/2$ \\
            & & & &           $\rightarrow \overline{\nu}u$ & $1/2$ \\
      \hline
      \hline
    \end{tabular}
    \caption {\small \label{tab:lqbrw}
               Leptoquark isospin families in the Buchm\"uller-R\"uckl-Wyler
               model. Charge conjugate processes are not shown.
	       For each leptoquark, the subscript denotes its weak isospin.
               For simplicity, the leptoquarks are conventionally indexed 
               with the chirality of the incoming {\em electron} which could 
               allow their production in $e^-p$.
               The variable $\beta_e$ denotes the branching ratio of 
               the LQ into $e+q$.
               }
\end{center}
\end{table*}
%
For LQs with $F=0$, the upper limits on the 
coupling obtained at $95 \%\,{\rm CL}$ are
shown as a function of the LQ mass in Figs.~\ref{fig:brw}a and b, 
for scalar and vector LQs respectively.
For masses above $\sim 270 \GeV$, these bounds improve by a factor
of up to $\sim 3$ the limits obtained in~\cite{H1LQ99} 
from the analysis of $e^+ p$ data at $\sqrt{s_{ep}} = 300 \GeV$.
Constraints corresponding to $F=2$ LQs are shown
in Figs.~\ref{fig:brw}c and d which extend those in~\cite{H1LQe-}
beyond the kinematic limit. 
For a coupling of electromagnetic strength $\alpha_{\rm em}$
($\lambda = \sqrt{4\pi\alpha_{\rm em}}=0.3$)
this analysis rules out LQ masses below $275$ to $325 \GeV$, depending
on the LQ type.

\begin{figure}[p]
\begin{center}
\begin{picture}(50,180)
\put(-30,60){\epsfig{figure=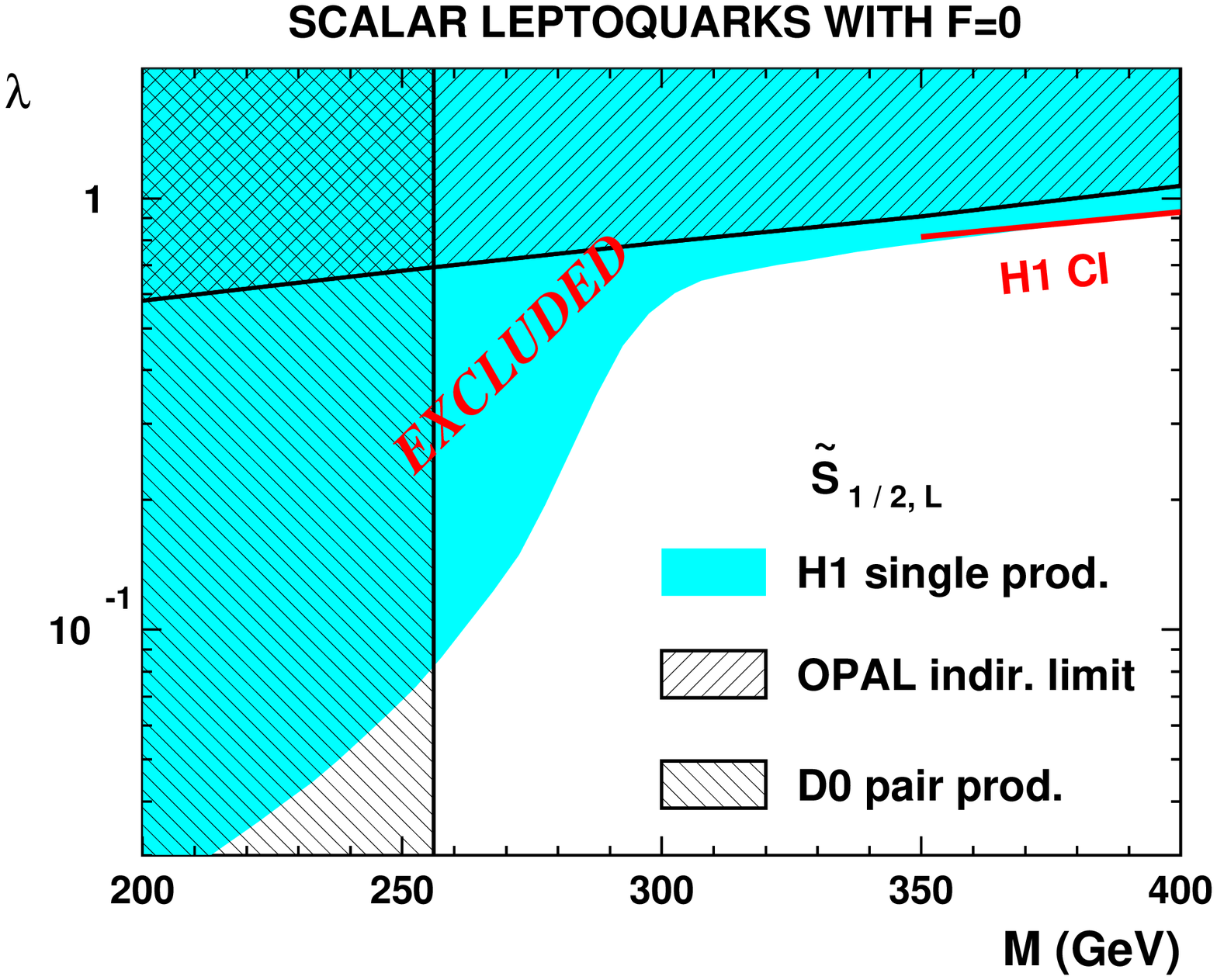,width=10.5cm}}
\put(-30,-30){\epsfig{figure=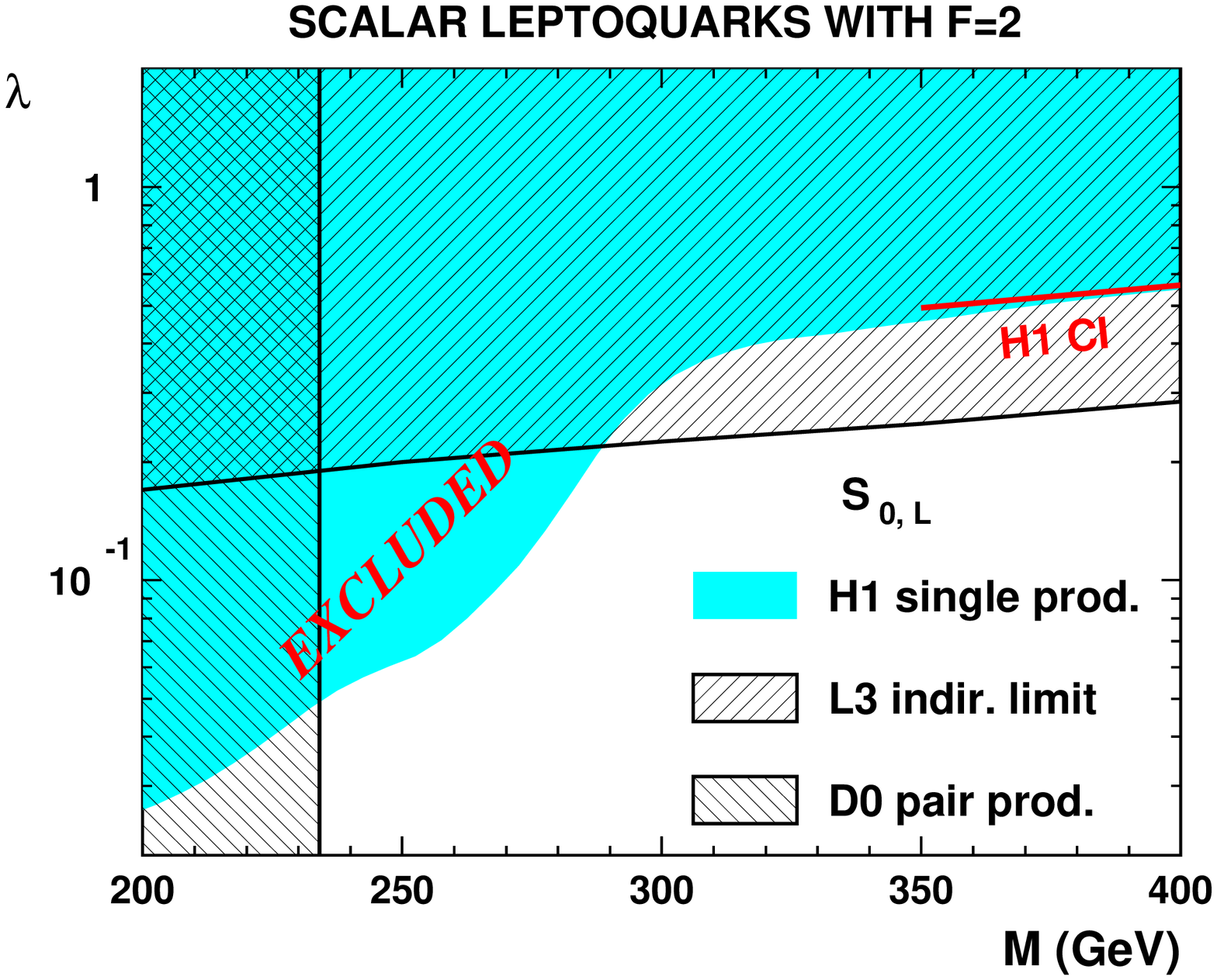,width=10.5cm}}
\end{picture}
\end{center}
  \caption{\label{fig:brwcompar}
  Exclusion limits at $95\%\,{\rm CL}$ on the coupling $\lambda$ as a
  function of the leptoquark (LQ) mass for $\tilde{S}_{1/2,L}$ (top) and
  $S_{0,L}$ (bottom) in the framework of the BRW model.
  The direct D0 limits 
  are independent of the coupling. For $\tilde{S}_{1/2,L}$ the indirect limit
  from OPAL is shown,
  whereas for $S_{0,L}$ the better indirect limit from L3 is shown.
  Constraints on LQs with masses above $350$\,GeV
  obtained from the H1 contact interaction (H1 CI) analysis~\cite{ci}
  are also shown, in the rightmost part of the figures.}
\end{figure}

Fig.~\ref{fig:brwcompar} summarises the constraints on the
$\tilde{S}_{1/2,L}$ and on the $S_{0,L}$ obtained by H1,
by the OPAL and L3 experiments at LEP~\cite{LQLEP}, 
and by the D0 experiment at the Tevatron~\cite{d0}.
The limits shown from LEP are from indirect constraints. 
The limits from the Tevatron are independent of the coupling $\lambda$ as
they were derived from the dominant pair production processes.
The H1 limits are comparable with those from the ZEUS experiment obtained
in a similar analysis~\cite{zeuslq}.
The limits at high mass values are also compared with those obtained
in a contact interaction analysis~\cite{ci}, which was
based on the measured single differential cross sections 
${\rm d}\sigma/{\rm d}Q^2$ from the NC process only~\cite{nccc03}.

\begin{figure}[p] 
\begin{center}
\begin{picture}(50,105)
\put(-30,-5){\epsfig{figure=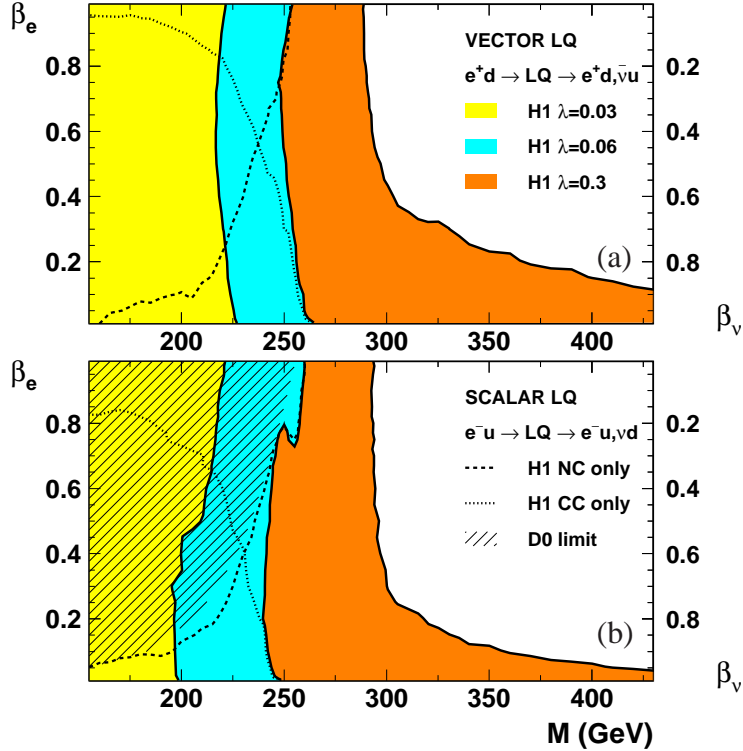,width=10cm}}
\put(50,60){(a)}
\put(50,10){(b)}
\end{picture}
\end{center}
  \caption{\label{fig:betaeplus}
  (a) Domains ruled out by the
  combination of the NC and CC analyses, for a vector LQ which couples to
  $e^+d$ (with the quantum numbers of the $V_{0,L}$) and decaying only into
  $eq$ and $\nu q$ for three values of the coupling $\lambda$.
  (b) Same as for (a) but for a scalar LQ coupling to $e^-u$ (with the
  quantum numbers of the $S_{0,L}$).
  The regions on the left of the full curves are excluded at $95\%\,{\rm CL}$.
  For $\lambda=0.06$, the part of the $\beta_e-M_{\rm LQ}$ 
  ($\beta_\nu-M_{\rm LQ}$) plane on the left of the dashed (dotted) curve 
  is excluded by the NC (CC) analysis alone. The branching ratios $\beta_e$ 
  and $\beta_\nu(=1-\beta_e)$ are shown on the left and right axes 
  respectively.
  The excluded domains cover $\beta_e$ values larger than $7.2 \times 10^{-5}$,
  $2.9 \times 10^{-4}$ and $7.1 \times 10^{-3}$ for $\lambda = 0.03$, $0.06$ 
  and $0.3$ respectively.
  In (b) the hatched region represents the domain excluded by the D0 
  experiment. The D0 limits do not depend on the value 
  of the coupling.}
\end{figure}

Beyond the BRW ansatz, generic LQ models can also be considered.
An example is provided by supersymmetric models\footnote{
   More general limits on squark production taking direct and
   indirect $R$-parity violating decay modes into account have been set 
   in~\cite{susylimit}.}
where the R-parity is violated by a $\lambda'_{1j1}$
($\lambda'_{11k}$) coupling, with the $\tilde{u}^j_L$ ($\tilde{d}^{k*}_R$)
squark having the same interactions with a lepton-quark pair as 
the $\tilde{S}_{1/2,L}$ ($S_{0,L}$).
In generic LQ models
other LQ decay modes are allowed such that the branching ratios $\beta_e$
and $\beta_\nu$ are free parameters.
Mass dependent constraints on the LQ branching ratios
can then be set for a given value of $\lambda$.
For a vector LQ coupling to $e^+ d$ (possessing the
quantum numbers of the $V_{0,L}$) and for $\lambda = 0.06$,
a domain of the $\beta_e$-$M$ ($\beta_{\nu}$-$M$) plane can be
excluded by the NC (CC) analysis as shown in Fig.~\ref{fig:betaeplus}a.
If the LQ decays into $e q$ or $\nu q$
only\footnote{It should be noted that $\beta_e + \beta_{\nu} = 1$ does
   not imply $\beta_e = \beta_{\nu}$ even when
   invariance under $SU(2)_L$ transformations is required.
   For example, when LQs belonging to a given isospin multiplet are not
   mass eigenstates, their mixing usually leads to different branching
   ratios in both channels for the physical LQ states. },
the combination of both channels rules out the part of the plane on 
the left of the second full curve from the left for $\lambda = 0.06$. 
The resulting combined bound is largely independent of the individual values
of $\beta_e$ and $\beta_{\nu}$.
Combined bounds are also shown for $\lambda=0.03$ and $\lambda=0.3$.
For a coupling $\lambda=0.3$ and high $\beta_{\nu}$ the limit extends to high
 mass values above the kinematic limit of resonant LQ production.
 For this part of the parameter space, the coupling
 $\lambda_{\nu}=\lambda\sqrt{\beta_\nu/\beta_e}$ is 
 large\footnote{In the BRW model, 
 $|\lambda_\nu|=|\lambda|$ for LQs coupling to both $eq$ and $\nu q$
 since $\beta_\nu=\beta_e$. 
 Here, in a generic LQ model, the effective coupling
 $\lambda_\nu$ at the LQ-$\nu$-$q$ vertex can be different from $\lambda$
 at the LQ-$e$-$q$ vertex.}
 but still satisfies
 $\lambda_{\nu}^2/4\pi<1$.
 A smooth transition is observed
 between limits driven by resonant production and limits driven by contact
 interactions.
Fig.~\ref{fig:betaeplus}b
shows similar exclusion limits as for Fig.~\ref{fig:betaeplus}a, 
for a scalar LQ possessing the quantum numbers of the
$S_{0,L}$ (which couples to $e^- u$).
The domain excluded by the D$0$ experiment at the Tevatron~\cite{d0} is 
also shown. For $\lambda$ greater than $\sim 0.06$, 
the H1 limits on scalar LQs extend considerably beyond the
region excluded by the D$0$ experiment.

To summarise, a search for leptoquarks
with fermion numbers $F=2$ and $F=0$
has been performed
using all $e^+p$ and $e^-p$ data collected by H1 between 1994 and 2000.
No signal has been observed and constraints on leptoquarks have been
set, which extend beyond the domains excluded by other experiments at
LEP and the Tevatron.
For a coupling of electromagnetic strength, leptoquark masses
below $275-325$ GeV, depending on the leptoquark type, can be ruled out.

\section*{Acknowledgements}

We are grateful to the HERA machine group whose outstanding
efforts have made this experiment possible.
We thank the engineers and technicians for their work in constructing 
and maintaining the H1 detector, our funding agencies for
financial support, the DESY technical staff for continual assistance
and the DESY directorate for support and for the
hospitality which they extend to the non DESY
members of the collaboration.

%

%

%

%

%

\end{document}